\documentclass[11pt]{article}
\usepackage[margin=1in]{geometry}
\usepackage{setspace}
\usepackage{array}
\usepackage{url}
\usepackage{tabularx}
\usepackage{tabulary}
\usepackage{multirow}
\usepackage{multicol}
\usepackage{url}
\usepackage{graphicx}
\usepackage{booktabs}
\usepackage{xcolor}
\definecolor{linkcolor}{HTML}{003399}
\definecolor{iconcolorcvwarning}{HTML}{F5BB17}
\definecolor{iconcolorcvcrimson}{HTML}{990000}
\usepackage[hypertexnames=false]{hyperref}
\hypersetup{
    colorlinks=true,
    breaklinks=true,
    citecolor=linkcolor,
    linkcolor=linkcolor,
    filecolor=magenta,
    urlcolor=linkcolor,
}
\usepackage[section]{placeins}

\usepackage{etoc}
\etocsetnexttocdepth{subsubsection}

\usepackage{amssymb}
\usepackage{longtable}

\usepackage{amsmath}
\usepackage{cleveref}
\begin{document}

%\begin{frontmatter}



\title{\emph{myAURA}: Personalized health library for epilepsy management via knowledge graph sparsification and visualization}


%
% Authors
%
\author{
    Rion Brattig Correia$^{1,2}$,
    Jordan C. Rozum$^{1}$,
    Leonard Cross$^{3}$, 
    Jack Felag$^{1}$,
    Michael Gallant$^{3}$,\\
    Ziqi Guo$^{1}$,
    Bruce W. Herr II$^{3}$, 
    Aehong Min$^{4}$, 
    Deborah Stungis Rocha$^{1}$, 
    Xuan Wang$^{3}$, \\
    Katy Börner$^{3,5}$,
    Wendy Miller$^{6}$,
    Luis M. Rocha$^{1,2}$
    \\
      \small $^{1}$ Department of Systems Science and Industrial Engineering, Binghamton University, Binghamton, NY, USA. \\
    \small $^{2}$ Instituto Gulbenkian de Ciência, Oeiras, Portugal \\
    \small $^{3}$ School of Informatics, Computing \& Engineering, Indiana University, Bloomington, IN, USA.\\
    \small $^{4}$ Donald Bren School of Information \& Computer Sciences, University of California, Irvine, CA, USA.\\
    \small $^{5}$ Alexander von Humboldt Fellow, Technische Universität Dresden, Dresden, Germany.\\
    \small $^{6}$ Indiana University School of Nursing, Indianapolis, IN, USA.\\
}

%%%
%%% TEMPLATE INFORMATION
%%%
% Word count: up to 4000 words.
% Structured abstract: up to 250 words.
% Tables: up to 4.
% Figures: up to 6.
% References: unlimited.

\date{}

\maketitle

\doublespacing

\begin{abstract} %no more than 250 words - Objective, Materials and Methods, Results, Discussion, and Conclusion

\textbf{Objective |}
We report the development of the patient-centered \textit{myAURA} application and suite of methods designed to aid epilepsy patients, caregivers, and researchers in making decisions about care and self- management.

\textbf{Materials and Methods |}
\textit{myAURA} rests on the federation of an unprecedented collection of heterogeneous data resources relevant to epilepsy, such as biomedical databases, social media, and electronic health records. A generalizable, open-source methodology was developed to compute a multi-layer knowledge graph linking all this heterogeneous data via the terms of a human-centered biomedical dictionary.

\textbf{Results |}
The power of the approach is first exemplified in the study of the drug-drug interaction phenomenon. Furthermore, we employ a novel network sparsification methodology using the metric backbone of weighted graphs, which reveals the most important edges for inference, recommendation, and visualization, such as pharmacology factors patients discuss on social media. The network sparsification approach also allows us to extract focused digital cohorts from social media whose discourse is more relevant to epilepsy or other biomedical problems. Finally, we present our patient-centered design and pilot-testing of \textit{myAURA}, including its user interface, based on focus groups and other stakeholder input.

\textbf{Discussion |}
The ability to search and explore \textit{myAURA}’s heterogeneous data sources via a sparsified multi-layer knowledge graph, as well as the combination of those layers in a single map, are useful features for integrating relevant information for epilepsy.

\textbf{Conclusion |}
Our stakeholder-driven, scalable approach to integrate traditional and non-traditional data sources, enables biomedical discovery and data-powered patient self-management in epilepsy, and  is generalizable to other chronic conditions.

\end{abstract}



{

%\begin{keyword}
}
%\end{keyword}

%\end{frontmatter}


%\section{Background and Significance}

%%
%% INTRODUCTION
%%
\section{Introduction}\label{sec:intro}

Epilepsy is a chronic neurological disorder that affects more than 3.4 million Americans and 65 million people worldwide \cite{EFA:2016:Epilepsy, ngugi2010estimation}. People with epilepsy (PWE) are at risk for lower quality of life, social isolation, depression, anxiety, medication-related symptoms, and premature death \cite{EFA:2016:Epilepsy, ngugi2010estimation,Hesdorffer:2013,Molleken:2010,Tomson:2016,Miller:2014:Dis,Austin:2012,England:2012,bazargan2020health,wood2022small}. 
Exacerbating these risks, PWE can wait up to 9 months to get a neurologist appointment and 6 months more to see an epileptologist, so many PWE are treated by general practitioners as they wait \cite{Hesdorffer:2013,miller2021evaluation,majersik2021shortage,elkhider2022predictors,ross2014option}.
Thus, alongside specialized medical care, self-management by PWE and their caregivers (PWEC) becomes essential for achieving desirable health outcomes \cite{Hesdorffer:2013}.
However, PWEC face uncertainty due to a daunting array of options about treatments, drugs, drug interactions and side effects, diet, lifestyle, and stigma.
Indeed, chronic health conditions unfold as a complex interplay among all these biological, psychological, and societal factors that change over time.
While much recent research has aimed to help patients retrieve health information online, the abundance PWEC typically discover from heterogeneous data sources makes it more difficult to distinguish the best treatment options available or even the relevance of information to an individual case.

Integrated and individualized information retrieval, as a personal health library, thus has a clear role to play in improving health outcomes for PWE and,
%
%This is particularly true for those who manage chronic conditions, such as PWE, 
%for whom there are currently no tools to integrate relevant online resources.
%
indeed, anyone with a chronic health condition. Qualitative and quantitative studies \cite{Austin:2012,Unger:2009, Miller:2014:Pat}, including those pursued under our project \cite{Min:2021,Min2023CHI}, show a clear %since information abundance makes it difficult to understand best treatment options or even relevance of information to each personal case.
need for visually engaging, easy-to-use, online tools for two key purposes:
    (i) to extract, classify, organize, and personalize information; and
    (ii) to provide automated recommendations in support of evidence-based decisions about treatment and self-management.
Despite the importance, there are currently no such online tools that integrate relevant information for PWEC.
They must conduct multiple separate searches of many different resources and manually comb through an array of often irrelevant and confusing results.

Here we present milestones of the ongoing \textit{myAURA} interdisciplinary project that aims to address this problem directly with data- and network-science methods to integrate multiple resources into a personalized, easy-to-use web service for PWEC.
To design this service according to their needs, our interdisciplinary team of experts in biomedical informatics, text and social media mining, visualization, user interface design, and epilepsy self-management work with patients, caregivers, and their advocates.
We also leverage a collaboration with important stakeholders at the Epilepsy Foundation of America (EFA), through an exclusive use agreement to obtain PWEC data from their website, discussion groups, and social media presence, and to recruit PWEC for our user study group and to provide general feedback about results.
All of this goes into computing a large-scale \textit{epilepsy knowledge graph}, comprised of a set of networks associating data from heterogeneous data sources relevant to PWEC. In addition, we discuss how computing its metric backbone, a network sparsification method based on removing edges that are redundant for shortest path computation \cite{Simas:2021}, yields a powerful method to infer, identify, visualize, and recommend personalized, relevant information for PWEC.

We also summarize our patient-centered methodology for designing a \emph{myAURA} application, with input from PWEC.
Per stakeholder needs and human-centered design specifications, when fully deployed, \emph{myAURA} will integrate practical, location- and patient-specific health-care information with targeted scientific literature, biomedical databases, social media platforms, and epilepsy-related websites with information about specialists, clinical trials, drugs, community resources, and chat rooms.
%
%It will build on innovative data and network science methods pursued in the research aims.
The innovative data- and network-science methods that \emph{myAURA} is designed upon drive the following three research aims:
%%%

\begin{enumerate}

\item  Produce a multi-layer epilepsy \textit{knowledge graph} by federating heterogeneous sources of large-scale data such as social media, electronic health records (EHR), patient discussion boards, scientific literature databases, and advocacy websites. %, and mobile app data.
This knowledge graph is built with epilepsy-focused terminology including the tagging of symptoms and medications.
%%%
\item  Develop recommendation and visualization algorithms based on automatically extracting the metric backbone of the knowledge graph, which, by reducing redundant edges, is likely to contain the information associations most relevant to a specific user's interests.
%For instance, recommendation algorithms based on the computation of shortest paths of the knowledge graph to extract backbones that are likely to contain relevant personalized information;
%%%
\item  Design and pilot test \emph{myAURA} 
%end-user studies to validate if and how the tool improves patient activation---the degree to which a person has the knowledge, skills, and confidence to manage epilepsy.
%
%This includes 
using focus groups studies that survey PWEC regarding their desired \emph{myAURA} content and its format, including interaction with mock-ups of the \emph{myAURA} interface to solicit suggestions for a more usable, valuable, and effective application.
\end{enumerate}

In summary, our immediate-goal as detailed in the following is to produce and visualize a knowledge graph representation of heterogeneous resources useful to PWEC.
The epilepsy knowledge graph supports a user-friendly web service to facilitate PWEC self-management and we also report on the interface design built from PWEC focus groups, as well as design requirements for other similar applications.
Our long-term goal is to generate a personal health library for PWEC and in so doing create a method that can be generalized to support self-management of other chronic diseases.

\section{Data and Methods}\label{sec:data-Methods-building-knowledge-graph}
\begin{figure*}[ht]
    \centerline{\includegraphics[width=\textwidth]{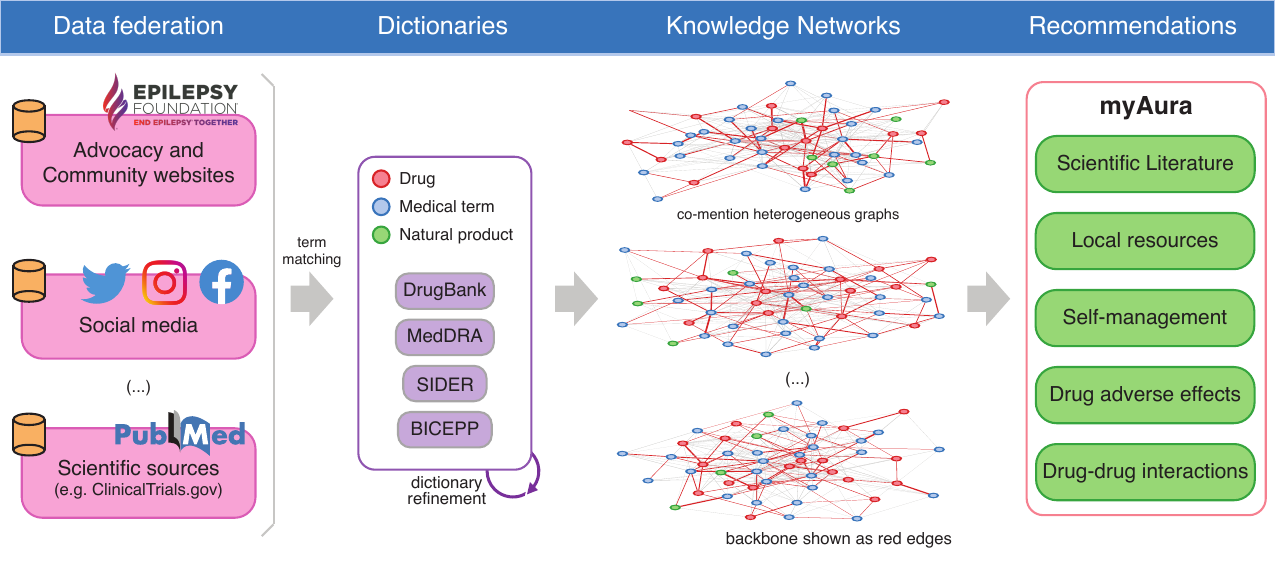}}
    \caption{
        Diagram of the overall \emph{myAURA} project:
        heterogeneous data federation;
        biomedical dictionary construction from various scientific resources;
        the constructed multi-layer knowledge network;
        the computed backbone of this epilepsy knowledge network; and
        the \emph{myAURA} application features from user-centered, such as scientific literature recommendations, local resource searches (e.g., epilepsy centers \& clinical trials), self-management support, and medication information
        .
    }
    \label{fig:myaura-diagram}
\end{figure*}

%A fundamental requirement for developing online tools is to integrate information resources relevant to the end-users.
%
A vital requirement for developing patient-centered tools is integrating information resources relevant to end-users.
To ensure that \emph{myAURA} meets the needs of PWEC, we not only engage them in its development, but also federate on their behalf data from epilepsy-specific advocacy and community websites, social media, EHR, research literature, and clinical and pharmacology databases.
We process the data from these resources to produce various large-scale knowledge networks \cite{Steyvers:2005, Johnson:1998} that are amenable to analysis with the powerful tools of network science \cite{Borner:2007:Net, Wasserman:1994, Monge:2003, Barabasi:2003} and machine learning \cite{Chakrabarti:2006, Borner:2011:Plu}.
 The overall architecture is depicted in Figure \ref{fig:myaura-diagram}.

%
% Data federation and pre-processing
%
\subsection{Data federation and processing}
\label{sec:data-federation}

In our federated database architecture we included two main groups of resources relevant to epilepsy and PWEC.

%%% Social Media (Twitter / Instagram / Facebook)
\subsubsection{Social media and community websites}
We have previously demonstrated the utility of social media data in the study of epilepsy and other biomedical problems \cite{wood2022small, Correia:2020:review, Correia:2016, Correia:2019:thesis,Min:2023,Wood:2017,wood2023time}, and here included \textit{digital cohorts} from Instagram, X (Twitter), Reddit, Facebook, YouTube comments, and the EFA website forums and Facebook discussion wall as detailed below.

\textit{Instagram} currently has more than 1.2 billion monthly active users.  
%
%It no longer allows access to its entire data stream (the Firehose), but it accepts requests for applications to query its data via an API, allowing the construction of hashtag- and timeline-based datasets. 
%
The current study uses a dataset collected between October 2010 and January 2016 via its API \cite{Correia:2016,Correia:2020:review}. 
This epilepsy-specific digital cohort contains 9,890 complete user timelines, i.e. all time-stamped posts of users who posted at least once about a drug used to treat epilepsy, resulting in a total of 8,496,124 posts.

From \textit{X} (\textit{Twitter}), using the historical gardenhose and the OSoMe data and tool set \cite{OSoMe:2016}, we collected  a random sample of 700,000 user timelines from from which we selected and processed  5,958 complete timelines with same criteria as for Instagram, containing 14,152,929 posts.

\textit{Reddit} is a user-moderated forum organized into over 100 thousand sub-forums called subreddits that are devoted to specialized topics. Reddit has over 57 million daily active unique user accounts and more than 13 billion posts and comments. Of particular interest is the r/Epilepsy subreddit, which is devoted to PWEC. This subreddit has been active since August 2010 and has more than 30 thousand unique users who have posted more than 277,367 comments/posts (typically larger than posts on Instagram or X).
For more direct comparison with \textit{Instagram} and \textit{X}, we identified a subcohort of $6,301$ users who posted at least once about a drug used to treat epilepsy. Their timelines contain a total of $219,459$ posts that, unless otherwise noted, comprise our epilepsy digital cohort for Reddit. 

\textit{YouTube}, used by an estimated 81 percent of Americans in 2021, is the most popular social media platform in the US. 
%
%It provides public access to an API for extracting video metadata, including closed captioning transcripts, video titles and descriptions, and user comments and replies.
%
Via its API, using the same criteria as for Instagram and X, we collected a digital cohort of more than 2 thousand users who have explicitly mentioned drugs used in the treatment of epilepsy, from a population of more than 330 thousand users who have engaged with epilepsy-related content over an 18-year period. 

From \textit{Facebook}, with support from the EFA and via a specially-developed application, we collected a small cohort of entire timelines of 12 victims of Sudden Unexpected Death in Epilepsy (SUDEP) for a study suggesting that SUDEP victims observe increased activity on Facebook prior to death  \cite{wood2022small}.

In summary, we collected over 48K complete user timelines, with over 23M posts, of X, Instagram, Reddit, YouTube and Facebook users who posted at least once about drugs related to epilepsy, or participated in forums on the topic (e.g., r/Epilepsy on Reddit).

%%% Mobile App and Web Search
%\textbf{Mobile App and web search}
%\emph{ChaCha} was a human-guided search engine providing free, real-time answers to any question, via its website, text messaging, and mobile applications \cite{ChaCha:2017}.
%\textbf{Chacha}. Short message service (SMS) and website messages from ChaCha are securely stored in a SQL database and cannot be shared in raw per copyright issues.
%
%We have the entire ChaCha dataset, which is of significant public health interest.
%
%We will also use Google Trends \cite{GoogleTrends} data curated for epilepsy-related terminology (see \cref{sec:dictionary}), as PI Rocha’s lab has done in other health-related tasks such as the study human reproductive patterns on a planetary scale [18].

%%% EFA website
%\textbf{Advocacy and community websites}.

While the social media sites provide large and broad platforms for investigating health-related signals, \textit{advocacy and community websites} provide data specific to epilepsy.
Via an exclusive use agreement, we have access to the EFA website (\url{epilepsy.com}), with more than 1 million unique user visits per month, and its highly used message boards, chat rooms, comment threads, and the MyEpilepsyDiary (which allows users of the EFA website to track medications, seizures, triggers, side effects, and symptoms).
Indeed, the social activity on the site is akin to those on social media \cite{Correia:2020:review} with the added research benefit that they are focused on the target PWEC community and their activities and health considerations.
Data was collected from 2004-2016, and it includes timelines of 22,938 active users with a total of 111,075 posts---the subcohort of users who posted at least once about a drug used to treat epilepsy is comprised of 8,488 user timelines with a total of 78,948 posts.
Additionally, we have been granted access to the EFA Facebook page, which has recently substituted the user forums (message boards and chat rooms) that were previously hosted on the EFA website. It has  ~115,000 followers with user comments from 2009-present.

\subsubsection{Biomedical and patient data} 

In addition to social media data, the federated database includes Clinical, pharmacological, health, and scientific databases, including EHR, which are relevant to epilepsy and PWEC and detailed below.

%%% Electronic Health Records
\textit{Electronic Health records}.
We use anonymized population-level EHR data extracted directly from health information systems.
It includes population-wide EHR from the public healthcare systems of the cities of Blumenau (Brazil, pop. 330.000) and Indianapolis (USA, pop. 864.447), and the whole of Catalonia (Spain, pop. 7,5M).
We curate these EHR data to compute knowledge graphs that uncover drug-drug interactions (DDI) and adverse drug effects (ADR) by risk level (major, medium, and minor), gender, and age. Analysis of these graphs, discussed below (\S \ref{sec:KG_DDI}),has already revealed important sex and age biases in all three populations \cite{Correia:2019:npj, sanchez2024prevalence}. 
Including these drug and symptom knowledge graphs in the \emph{myAURA} data federation allows us to focus on epilepsy-relevant DDI and ADR, as well as epilepsy-specific biases. 
Moreover, these graphs enable the future comparison and analysis of drug interactions, adverse reactions, symptoms, and temporal comorbidity trajectories in \emph{myAURA}’s user population,  with those observed in these independent, larger patient populations. This will facilitate issuing medication and symptom warnings to \emph{myAURA} users and PWEC at large \cite{sanchez2024prevalence}.

%In our design, the direct upload of EHR data by \emph{myAURA} users is also considered. This may include, but is not limited to, files that patients acquire directly from their medical providers. The files are parsed to populate demographics and date-stamp information about drug prescriptions, medical appointments, ER visits, medical exams, and so on.
%
%While our user focus group has already agreed to donate this sensitive data, in general it will be ingested into our federation only a final deployment of the service.
%
%For development and testing purposes, we use epilepsy-related data from the public healthcare systems as described above.

%\underline{Scientific literature}.
%
%%% PubMed/MEDLINE
\textit{PubMed} is a service of the National Library of Medicine, a ``free resource supporting the search and retrieval of biomedical and life sciences literature with the aim of improving health–both globally and personally'' that includes over 35 million citations dating back to the 1860s.
We process updated local copies of the entire PubMed/MEDLINE database (28 million citations) and use them in the \emph{myAURA} knowledge graph to enable the recommendation of scientific literature relevant to PWEC (e.g., abstracts, MeSH terms, and references related to medications).

%%% Clinical and pharmacology databases and websites
%\textit{Clinical and pharmacology databases}.
\textit{ClinicalTrials.gov} is a central registration site for clinical trials operated by the National Institutes of Health that has been available to the public since 2000 \cite{Zarin:2011}.
%
%It was established by law in 1997 as a resource for patients, their family, health care professionals and researchers, to provide easy access to information on clinical studies.
%
Both publicly and privately funded trials are represented.
The full dataset is available online and we ingested a local copy of the data into the Scholarly Database at Indiana University, which has been integrated into myAURA’s federated database and tagged with its dictionary for knowledge graph construction. As explained below (\S \ref{sec:dictionary}) , the dictionary construction required processing several pharmacology and clinical resources, such as \textit{DrugBank} \cite{DrugBank:v5}, SIDER \cite{SIDER}, FAERS \cite{FAERS}, \textit{MedWatch} \cite{Medwatch}, and \textit{Drugs.com}, to link relevant pharmacology and symptom information.
%

%%% Other Resources
%\underline{Other useful resources}.
%
Via our PWEC user focus groups (\S \ref{sec:user-interface}), we identified and ingested other resources deemed most useful to patient experiences,
%
%and have validated with our focus groups is 
%
such as the American Epilepsy Society’s \textit{Find a Doctor }Database \cite{FindADoctor}, which contains geographic locations of all epileptologists in the United States.
%
%we validated our inclusion 
%
Furthermore, based on the focus group study, the design of \emph{myAURA}'s user interface (\S \ref{sec:user-interface}) includes local transportation information, integrating services such taxi, Lyft, UBER, and other public transportation via their APIs or Google Maps.

\subsection{Biomedical dictionaries \& sentiment analysis}
\label{sec:dictionary}

To build an epilepsy knowledge graph, relevant concepts are needed to define its nodes.
Indeed, a key aspect of federating the various resources is the construction of dictionaries with all the relevant entities and terminology for automatically tagging text of potential relevance to epilepsy. 
As others have done for studying depression using Twitter \cite{Sarker:2015:Uti, Schwartz:2016}, we first included terms obtained from clinicians and extracted from epilepsy patient social media \cite{Correia:2016}.
This was supplemented with additional dictionaries that were previously carefully curated by pharmacology and biomedical informatics experts to be used in pipelines for extracting DDI  from the scientific literature \cite{Kolchinsky:2013, Kolchinsky:2015:Ext, Correia:2016, Correia:2019:thesis, zhang2022translational}.
These dictionaries contain +170,000 standardized terms from sources such as FDA drug labels, DrugBank \cite{DrugBank:v5}, SIDER \cite{SIDER}, BICEPP \cite{LinFPY:2011},  FAERS \cite{FAERS}, and a standardized medical terminology dictionary built from clinical notes, MedDRA\cite{MedDRA}.
Parent terms and synonyms were associated in a hierarchical manner, e.g., Prozac is resolved to fluoxetine and cold to nasopharyngitis.
Drug terms were expert-curated to best capture experimental evidence for DDIs and adverse drug reactions in the literature. Our \textit{myAURA} dictionary was shared with the community with a corpus of PubMed articles and sentences with direct experimental evidence of DDI \cite{zhang2022translational,wu2020translational}.

Clinical terminology is not tailored for social media language so it can bias biomedical inference pipelines,  such as the pharmacological surveillance that we reviewed  elsewhere \cite{Correia:2020:review}. 
We therefore refined the dictionary via human-centered curation, with 14 annotators, iterative design of annotation guidelines, and the \textit{Instagram} epilepsy digital cohort collected as described above. 
We showed that text mining pipelines built for scientific literature must be refined when applied to general-purpose social media text where many topics are discussed simultaneously.
Indeed, the removal of just the 12 terms deemed most ambiguous by human annotators reduced false-positive tokens in the tagging of social media data, and significantly improved recommendations on the epilepsy knowledge graph \cite{Min:2023}. 

Using the final \textit{myAURA} dictionary, all textual data from the federated resources above were tagged with its terms, to ensure capturing discourse that is most relevant for PWEC such as epilepsy symptoms and terminology, drugs and pharmacology, natural products, and adverse reaction terminology.
Specifically, we extracted and tagged the relevant text fields from all the federated resources, such as social media posts, prescription data e EHR, or eligibility criteria in clinical trials. These were then indexed in a data warehousing system for easy linking of relevant concepts to text units, users and all data fields in the federated data resources above.
The tagged concepts/terms are subsequently used as nodes in the epilepsy knowledge graph described below (\S \ref{sec:knowledge-graphs}).
%
%These datasets were loaded onto relational- (MySQL) and document-based (Mongo) databases for fast access.

%Going forward, we will tag and time-stamp EHR from \emph{myAURA} users who choose to upload them, thus effectively processing EHR as dictionary-tagged patient health trajectories.
%

In addition to the biomedical-dictionary tagging, we used several dictionary-based sentiment analysis tools such as ANEW \cite{Bradley:1999}, VADER \cite{Hutto:2014}, and LIWC \cite{Tausczik:2010} 
to tag each post, tweet, and comment in social media and community website data sources, with a mood state along sentiment dimensions including valence (happy/sad), arousal (calm/excited), and dominance (in-control/dominated) \cite{Wood:2017}.
This allows us to estimate individual and collective  psychological mood state of the epilepsy digital cohorts, affording various types of health-related discoveries, as we reviewed for the biomedical data science community \cite{Correia:2020:review}.
For instance, as part of this project, we studied a small Facebook cohort of victims of SUDEP and showed that certain sentiment measures such as increased or altered verbosity may be predictive of this serious outcome, an important result for stakeholders
%
%in a small cohort of SUDEP patients that these sentiment measures extracted from Facebook timelines may play an important role in characterizing patient outcomes
\cite{wood2022small}.

\subsection{Building the \textit{myAURA} epilepsy Knowledge graph}
\label{sec:knowledge-graphs}
Given that the textual items of the federated data resources were tagged with dictionary terms, it is straightforward to build weighted graphs (i.e., networks), where edges denote a co-occurrence proximity measure (or its inverse, distance), e.g., the co-occurrence of drugs and medical terms on social media posts or EHR entries. 
Specifically, given the set $X$ of all terms, we first compute a symmetric co-occurrence matrix, $R_w(X)$, whose entries $r_{xy}$ denote the number of textual units $w$ where terms $x$ and $y$ co-occur \cite{Simas:2015,Simas:2021}.
Unit $w$ may denote a PubMed abstract \cite{Abi-Haidar:2008}, a user timeline-window on Instagram \cite{Correia:2016,Correia:2020:review}, or an EHR prescription period \cite{Correia:2019:npj,sanchez2024prevalence}.
The diagonal entries of this matrix, $r_{xx}$, denote the total number of times term $x$ was mentioned in a unit of analysis with any other term in the dictionary $X$: $r_{xx} = \sum_{y \in X : y \neq x} r_{xy}$.

To measure a normalized strength of association among the $X$ terms, we compute a \textit{proximity graph} $P(X)$ whose edge weights are given by the Jaccard similarity \cite{Jaccard:1901, Grefenstette:1994} (though other measures are possible \cite{Simas:2015,Ciampaglia:2015,Baeza-Yates:1999,Turney:2001,Klir:2005}): 

\begin{equation}
    \label{eq:proximity}
    p_{xy} = \frac{r_{xy}}{r_{xx} + r_{yy} - r_{xy}}, %\quad \forall {x_i,x_j} \in X,
\end{equation}

\noindent where $p_{xy} \in [0,1]$ denotes a \textit{proximity} between two terms $x$ and $y$. When the terms never co-occur on textual units $w$ we have $p_{xy}=0$, and when they are both always mentioned together we have $p_{xy}=1$; naturally, $p_{xx}=1$.

These proximity or associative \textit{knowledge graphs} (KG) are simple, yet powerful, data representations to study the relationships among different entities.
We have used these associative \textit{knowledge graphs} (KG) to build competitive recommender algorithms \cite{Simas:2015,Rocha:2002:Sem,Rocha:2005,Simas:2012}, biomedical text mining pipelines \cite{Abi-Haidar:2008,Verspoor:2005,Kolchinsky:2010,Lourenco:2011}, scientific maps \cite{manz2022viv,Borner:2021,azoulay2018toward,Borner:2015:Atlas,ginda2023introducing}, network inference in biomedicine \cite{sanchez2024prevalence,correia2024conserved,Correia:2019:npj,Correia:2016,Simas:2021}, and automatic fact-checking \cite{Ciampaglia:2015}.
Also, computing KGs is scalable; it depends only on pairwise comparison of vectors for each pair $(x,y)$ \cite{Simas:2015,Kalavri:2016}.

Finally, the \textit{myAURA} epilepsy KG,  $\varepsilon = \{P^s (X)\}$, is composed of all the KGs $P^s (X)$ for each federated data source $s$  described above and dictionary terms in $X$. 
Because the dictionary terms, $x$, are shared across KGs, this can be cast as a multi-layer graph where term associations for each data source are represented separately, with inter-layer edges connecting the same dictionary terms on each layer,
as represented in Figure \ref{fig:myaura-diagram}.

%%
%% RESULTS
%%
\section{Results}
\subsection{Studying Drug-Drug Interaction via KGs}
\label{sec:KG_DDI}

\begin{figure*}[ht]
    \centerline{\includegraphics[width=\textwidth]{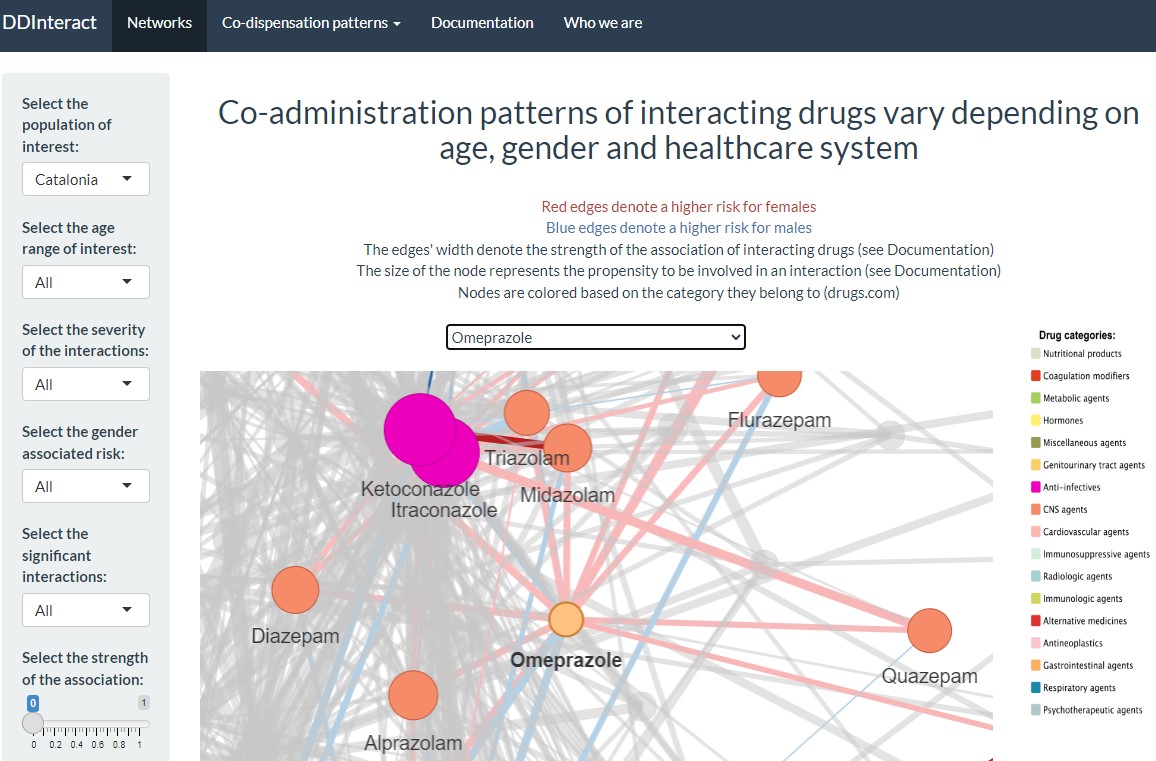}}
    \caption{
        Screenshot of the DDIInteract tool \cite{Sanchez-Valle_DDIInteract,sanchez2024prevalence} displaying proximity graphs of drug-drug interaction networks extracted from EHR, highlighting the most frequent interactions associated with Omeprazole.
    }
    \label{fig:DDIInteractTool}
\end{figure*}

Let us exemplify the utility of the\textit{ myAURA} KG with the networks obtained from EHR data.
We processed 18 months of EHR data from all 133K patients in the primary care public health system of Blumenau, Brazil, then conducted a large longitudinal study of the prevalence of known DDIs. We discovered very significant sex and age biases, even after correcting for multiple factors. Women and older patients were at significantly higher risk of being exposed to DDI than their polypharmacy regimens would suggest \cite{Correia:2017:CCS}, with several drugs used in epilepsy implicated (e.g. see Figure \ref{fig:DDIInteractTool}). To understand if such biases occur beyond Brazil, we showed generalizability and reusability of our pipeline using additional primary care data from distinct locations: Indiana State, 2 yrs, 265K patients; and Catalonia, Spain, 10 yrs, 5.5M patients. We found very similar sex and age biases in the prevalence of known DDI in both locations, albeit sometimes involving distinct drugs \cite{sanchez2024prevalence}. The analysis also revealed actionable interventions that easily reduce both biases and the burden of DDI, e.g., by replacing the drug Omeprazole with other proton-pump inhibitors. 
This study was enabled by the EHR data federated with our dictionary and other databases, and the KGs built from them to characterize drug interactions in proximity graphs for each population.  Indeed, a web tool (DDIInteract \cite{Sanchez-Valle_DDIInteract,sanchez2024prevalence})  was built by external collaborators to enable third-party analysis of the DDI KGs from all the \textit{myAURA} EHR datasets (Figure \ref{fig:DDIInteractTool}), further demonstrating reusability of our data and KGs.

The focus group user studies discussed below (\S \ref{sec:user-interface}) revealed that detailed pharmacological information is of particular importance to epilepsy patients. Therefore, in addition to studying DDI in EHR data, we also built text classifiers to identify PubMed abstracts (and sentences) with direct experimental evidence of DDI. We trained/fine-tuned classical and large language models like BioBERT and ChatGPT on the human annotated DDI corpora that we helped develop \cite{zhang2022translational} and on the refined dictionary described above, and they performed very well (MCC $\approx 0.9$ for in vitro and in vivo, and $\approx 0.8$ for clinical evidence) \cite{wang:2024}. The recommendation of relevant experimental DDI evidence supports functionalities in the \textit{myAURA} mockup  discussed below (\S \ref{sec:user-interface}), e.g., when users click on nodes representing drugs in the KG visualizations. This study also revealed knowledge gaps in the scientific literature by identifying drug pairs in need of experimental in vitro, in vivo, or clinical DDI studies \cite{wang:2024}. 

%%% Source code and Software
%\textbf{Source code and software}.
%

We developed \textit{standardized open-source code} for efficient data ingestion, preprocessing, dictionary term matching, construction, storage, and joining of networks from different data sources into \textit{PostgreSQL}, and for computing the metric backbone discussed next. It provides a unique, fast, and streamlined process that, in a few commands, reproduces the complete KG construction pipeline for each independent data source that gets updated. 
This source code is publicly shared through github\footnote{at \href{https://github.com/cns-iu/myaura}{github.com/cns-iu/myaura} for KG construction and \href{https://github.com/CASCI-lab/}{github.com/CASCI-lab/} for the backbone extraction pipeline.}, the most widely used public repository of public software, with appropriate licenses that allow other researchers to re-use and build on our source code.

\subsection{The Metric Backbone for KG sparsification}
\label{sec:KG_backbone}

As discussed in the previous subsection, KGs are multidimensional representations useful for inference and interpretation of biomedical issues from large scale data.
However, they are often dense with many edges that are not relevant for analysis, inference, link prediction, and recommendation---and furthermore impair visualization and slow down computation.
Thus, we developed a sparsification methods to facilitate analysis and visualization of \textit{ myAURA}'s epilepsy KG and other biomedical informatics problems based on network data.

Many network inference methods depend on shortest paths, which are computed on \textit{distance graphs} $D(X)$ easily obtained from the proximity graphs of the epilepsy KG $\varepsilon = \{P^s (X)\}$ via the nonlinear map $\varphi$ applied to the weights of the latter (given by eq. \ref{eq:proximity}):

\begin{equation}
    \label{eq:distance}
    d_{xy} = \varphi(p_{xy}) = \frac{1}{p_{xy}} -1, %\quad \forall {x_i,x_j} \in X, 
\end{equation}

\noindent where $d_{xy} \in [0, +\infty]$, $d_{xx}=0$ and the resulting distance weights are symmetrical and inversely proportional to the strength of association between terms; i.e. they convey a measure of distance necessary to compute path \textit{length.}

Shortest paths allow us to infer the strength of indirect association (or likely transmission): If $x$ is connected to $z$ with a finite distance, and $y$ is similarly connected to $z$, the length of the shortest indirect path quantifies how close $x$ is to $y$, or the ``cost'' of transmitting information from $x$ to $y$ via $z$. This type of inference is ubiquitous in network problems \cite{Borner:2007:Net, Wasserman:1994,Monge:2003, Barabasi:2003}, including link prediction \cite{LuLinyuan:2011} and recommendation \cite{LiXin:2013, DongYu:2012}, our inference of DDI from social media and EHR \cite{sanchez2024prevalence,Correia:2016,Correia:2019:npj,Correia:2020:review}, automated fact-checking on Wikipedia \cite{Ciampaglia:2015}, and epidemics on social networks \cite{Correia:2023contact,soriano2023semi}.

\begin{figure*}[ht]
    \centerline{\includegraphics[width=1\textwidth]{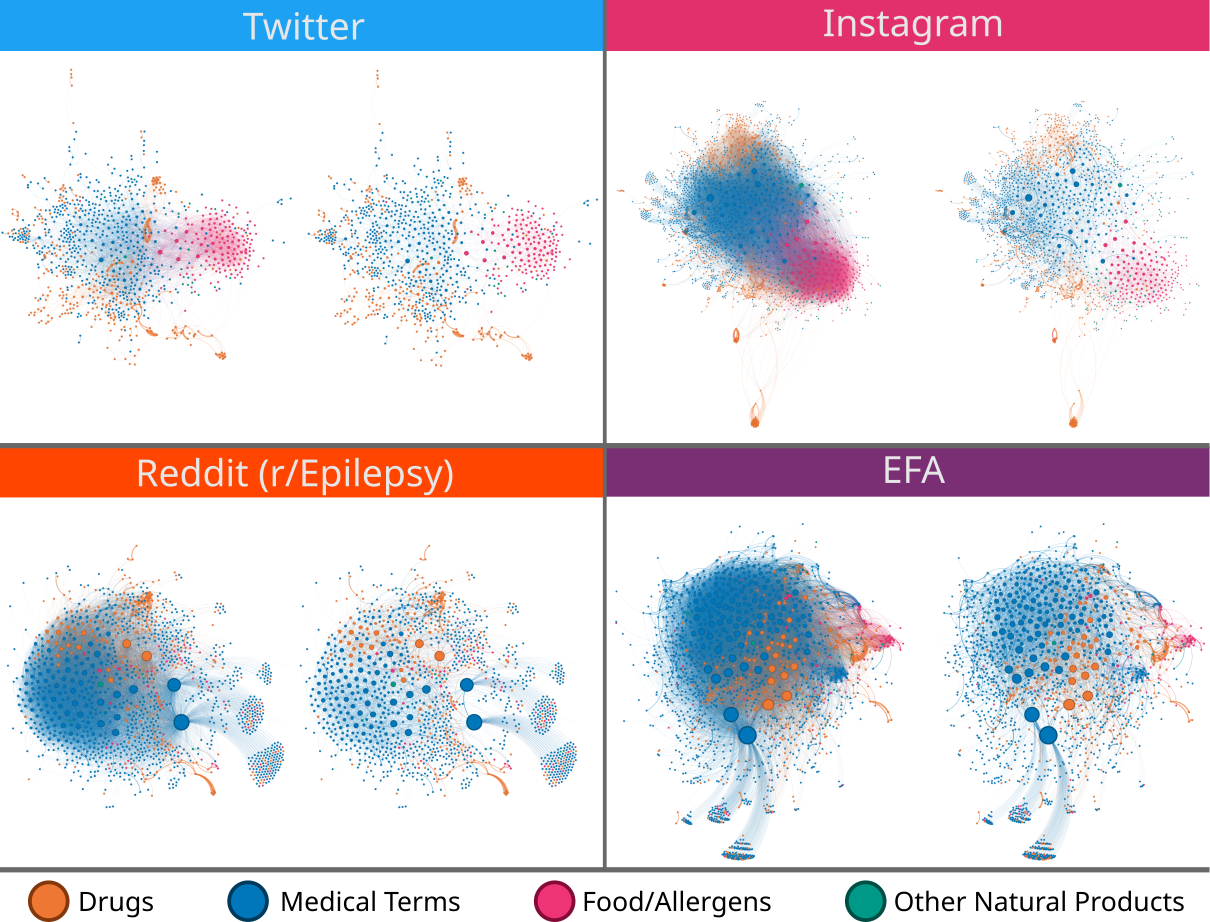}}
    \caption[]{
        Original distance graph (left) and its metric backbone subgraph (right) for each of the social media layers of \textit{myAURA}'s multigraph KG. Nodes are sized according to (unweighted) degree in original network; Reddit and EFA networks computed only from the drug mention subcohorts, to better compare with Twitter and Instagram networks. Node positions are determined by the \textit{Force2Atlas} method applied to the original networks. Only the largest component of each network is shown (the backbone preserves network components).
    }
    \label{fig:SocialMediaKGs}
\end{figure*}

We have shown that such distance graphs obtained from real-world data are typically not metric, but rather \textit{semi-metric} \cite{Simas:2015,Rocha:2002:Sem}: The \textit{triangle inequality} ($d_{xy} \leq d_{xz} + d_{zy}$) is not observed for every edge of $D(X)$ \cite{Galvin:1991}. That is, the shortest distance between at least two nodes in the graph is not the direct edge, but rather an indirect path via other nodes. 

Computing shortest paths of a distance graph, where path length is the sum of constituent edge (distance) weights ($d_{xy} = d_{xz} + d_{zy}$), e.g., via Dijkstra’s algorithm \cite{Dijkstra:1959}, yields its \textit{metric closure} $D^{C}(X)$, a new graph that obeys the triangle inequality at every edge \cite{Simas:2015}. If an edge in the original graph is semi-metric, its weight gets replaced by the length of the shortest indirect path between the nodes it connects. 
In other words, the metric closure (or All Pairs Shortest Path Problem \cite{zwick2002all}) is the graph obtained by computing the shortest paths between all pairs of nodes in the distance graph and replacing the original distance edges $d_{xy}$ with the length of the shortest path between $x$ and $y$: $d^{C}_{xy} =
%\ell_{xy} = 
d_{xz_1} + d_{z_1z_2} + \ldots + d_{z_\delta y}$, via an arbitrary number $\delta$ of intermediary nodes $z_k$.
In summary,  edge weights $d_{xy}$ of $D(X)$ that do not change after computation of the metric closure $D^{C}(X)$ are\textit{metric} because they obey the triangle inequality---there is no indirect path shorter than the direct edge between $x$ and $y$---while those that change, are the semi-metric edges.

Significantly, there is a \textit{metric backbone} subgraph $D_b (X)$ \cite{Simas:2021} of the original graph $D(X)$ that is invariant under the metric closure and is sufficient to compute all shortest paths: $D_b^C (X) \equiv D^C (X)$. 
The edge weights of the metric backbone graph are given by:

\begin{equation}
    \label{eq:backbone}
    b_{xy} = \begin{cases} d_{xy},& \mbox{if } d_{xy}=d^{C}_{xy} \\ +\infty,& \mbox{if } d_{xy} > d^{C}_{xy} \end{cases}\, , % \forall x_i,x_j \in X,
\end{equation}

\noindent where $b_{xy} = +\infty $ means there is no direct edge between $x$ and $y$ in the distance backbone graph.

The size of the backbone subgraph, in relation to the size of the original  graph, defines the amount of \textit{redundancy} in the network. Edges not on this backbone are superfluous in the computation of shortest paths and in all network measures derived from shortest paths (e.g., betweenness centrality). 
Importantly, the metric backbone is an algebraically principled network sparsification method with unique features: it (a) preserves all connectivity and shortest-path distribution, (b) does not alter edge weights or delete nodes, (c) is exact, not sampled or estimated, and (d) requires no parameters or null model estimation \cite{Simas:2021}. Furthermore, it outperforms available state-of-the-art network sparsification methods in (e) preserving the community structure of the original graph \cite{Correia:2023contact}  and (f) recovering most of the original (macro and micro) transmission dynamics in social contact networks, while revealing the most important infection pathways in epidemics, and resulting in greater reduction without breaking apart the original network \cite{correia2024conserved,soriano2023semi,Simas:2021}.

\begin{table}[]
    \centering
    \caption{
        Statistics of layers of the myAURA KG originating from various data sources.
        Columns show the number of nodes, edges, and the size of metric backbone as the proportion of edges kept.
        Data for EFA and r/Epilpsey refer to the subcohorts of users who posted at least once about a drug used to treat epilepsy.
    }
    \label{tab:knowledge-graphs}
\begin{tabular}{@{}lllll@{}}
    \toprule
    KG Network                & Nodes & Edges & \% metric \\
    \midrule
    PubMed*                    & 8,891 & 590,781 & 18.59\% \\
     Clinical Trials             & 1,275 & 31,371 & 53.75\% \\

    Instagram                  & 1,686 & 25,235 & 15.1\% \\
    X (Twitter)                     & 1,022 & 5,082 & 37.0\% \\
    r/Epilepsy (Reddit)         & 1,270 & 17,558 & 17.0\% \\
        EFA                        & 1,529 & 33,795 & 15.7\% \\
    \bottomrule
    \multicolumn{3}{l}{\footnotesize*only epilepsy related publications.}
\end{tabular}
\end{table}

All layers of \textit{myAURA}'s KG have a small backbone (large amount of redundancy) as can be seen in Table \ref{tab:knowledge-graphs}) and Figure \ref{fig:SocialMediaKGs}.
This is coherent with what is observed in networks across biological, technological, and social domains, which typically possess very small metric backbones---revealing that network robustness to attacks and failures seems to stem from surprisingly vast amounts of (shortest-path) redundancy \cite{Simas:2021}. For instance, the metric backbone of the KG of more than 3 million concepts extracted from Wikipedia is composed of only 2\% of the original edges, but it is sufficient to compute all shortest paths used by our automated fact-checking inferences \cite{Ciampaglia:2015}. Likewise, the metric backbone of a protein interaction network of more than 11K human genes involved in spermatogenesis comprises $\approx 10\%$ of the original edges \cite{correia2024conserved}. The 90\% of edges not on the backbone were also obtained from experimental evidence, but they are redundant for shortest paths and likely less important for regulatory pathways, which led to our discovery of new genes involved in male infertility \cite{correia2024conserved}. Similarly, the backbones of social contact networks important for epidemic spread are within 5-20\% \cite{Correia:2023contact,Simas:2021}; those of the human brain connectome and functional and multiomic gene co-expression networks are typically 5-11\% \cite{Simas:2021} and have distinguishing network features that enable effective classification between healthy and diseased human cohorts in Alzheimer’s, autism, depression, and psychotic disorder \cite{dorsant2023gene, Simas:2016, Peeters:2015}.

These observations show that the metric backbone is more than a mathematical construct and ``has a phenotype'': its measurement in many biomedical and social complexity problems reveals important functional characteristics, such as community structure, information spreading dynamics, and the most important (central) network nodes, edges, and pathways for inference \cite{Simas:2021,Correia:2023contact,soriano2023semi}. 
Additionally, since the backbones of large networks are typically very small, this natural sparsification provides substantial memory and computational parsimony in storing and analyzing them \cite{Kalavri:2016}.
Certainly, removing edges that are redundant for shortest paths yields a powerful sparsification methodology that facilitates analysis and visualization of KGs \cite{Simas:2021}, which we illustrate next (\S \ref{sec:backbone_inference}).

An open-source Python package for metric backbone extraction and analysis, \textit{DistanceClosure} \cite{DistanceClosure}, was developed to be compatible with \textit{NetworkX}, for interoperability with common graph formats (e.g. \textit{GraphM}L, \textit{GML}).

%
% The Backbones
%
%\section{The backbone of a knowledge graph}
%\label{sec:backbone}

\begin{figure*}[ht!]
    \centerline{\includegraphics[width=\textwidth]{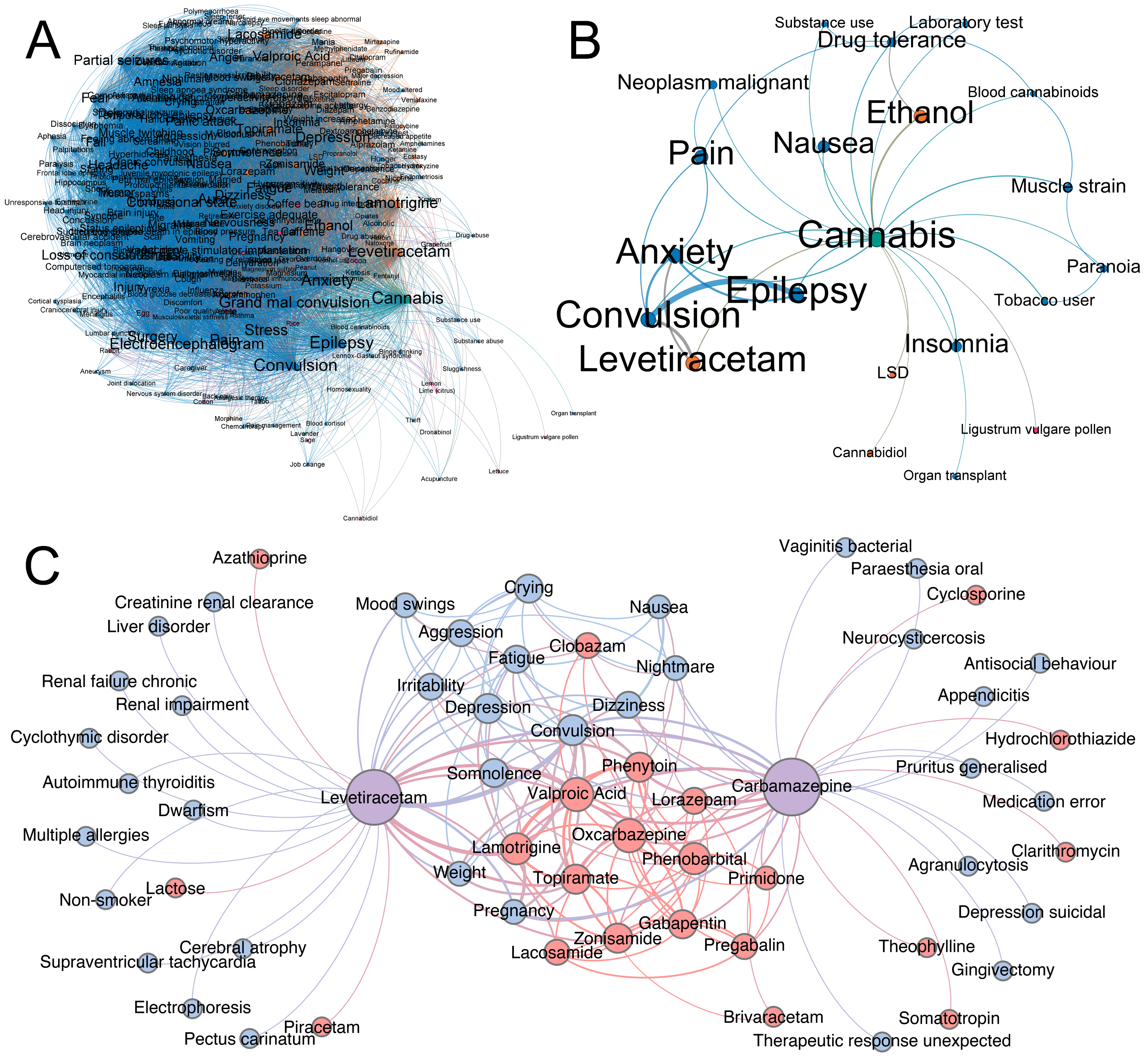}}
    \caption[]{\textbf{A:} Ego network for the term ``Cannabis'' using a Force2Atlas layout. \textbf{B.} The same network after sparsification via the metric backbone using a manual layout.
    Ego-networks in A and B are subgraphs of the full and backbone \textit{Reddit} KGs depicted in Figure \ref{fig:SocialMediaKGs}, respectively.
     \textbf{C:}   Subgraph of the backbone of the EFA KG depicted in Figure \ref{fig:SocialMediaKGs}.
    Terms directly connected, on the backbone, to both \textit{Levetiracetam} and \textit{Carbamazepine}, two drugs commonly prescribed to treat epilepsy.
        Note, some terms are shared by the two drugs (such as drugs commonly co-prescribed), while others are specific to each drug separately.
        Node colors denote dictionary term type: medical term (blue), drug (red), or queried term (purple).
    }
    \label{fig:backbone-graph}
\end{figure*}

\subsection{Analysis of myAURA's KG backbones}
\label{sec:backbone_inference}

%NOTE: Also use bit from proposal about the filtering method.

KG sparsification  enables various types of inference, extraction and recommendation from digital libraries, automatic fact-checking to protein-protein interaction extraction \cite{Simas:2021,Ciampaglia:2015,Rocha:2005,Rocha:2002:Sem,Simas:2012,Abi-Haidar:2008,Verspoor:2005,Kolchinsky:2010,Lourenco:2011}.  
Let us exemplify with the extraction of focused digital cohorts from social media, which are most useful to study the interplay between human behavior and medical treatment in chronic diseases such as epilepsy \cite{wood2022small,Correia:2020:review}.

Social media sites vary in the generality of their discourse; while \textit{X} and \textit{Instagram} simultaneously trade in a wide range of topics, \textit{Reddit} subgroups and the EFA discussion forums are much more focused on health-related discourse. The metric backbones of the \textit{myAURA} KGs from \textit{Instagram}, \textit{X}, \textit{r/Epilepsy}, and EFA forums are similar in size: $\approx 16\%$ of original, except X with $37\%$ (see Table \ref{tab:knowledge-graphs} and Figure \ref{fig:SocialMediaKGs}).
Even though user timelines were harvested with the same criterion (at least one post mentioning an epilepsy drug) in all platforms, the proportion of users who contribute to backbones (users with at least one post containing a pair of dictionary terms represented by an edge on the backbone) is quite distinct. A much higher proportion of users contribute to the backbone in epilepsy-focused than in general-purpose social media: 65 and 71\% on \textit{Instagram} and \textit{X} vs. 95 and 93\% on EFA forums and\textit{ r/Epilepsy}---as showed in detail in \cite{guo:2024}. In other words, in the general-purpose platforms there are a lot more users who do not contribute to any shortest-path inference on the derived KGs; they contribute to redundant KG edges. 

Interestingly, there is a clear discourse distinction between users who contribute to the backbone from those who do not. Using the human-annotated corpus of \textit{Instagram} posts utilized to refine the \textit{myAURA} dictionary (see \S \ref{sec:dictionary}) \cite{Min:2023}, we observed that the false positive rate (dictionary terms used without medical relevance) is significantly higher for the set of users who do not contribute to the backbone (32\%) than for those who do (14\%). Moreover, this difference is not a matter of engagement because false positive rates are similar for users who post a lot (13\%) or little (18\%). In sum, this backbone filtering methodology enables the extraction of focused digital cohorts from general-purpose social media, such as \textit{X} or \textit{Instagram}, by identifying user sets more like those on special-purpose forums of biomedical relevance such as the EFA and \textit{r/Epilepsy}. In other words, the metric backbone sparsifcation of KGs can be used to increase  personalization of social media data for a specific health problem \cite{guo:2024}.

The sparsification of original networks into their metric backbone subgraph can also be used to uncover drug side effects while highlighting key drug and medical term associations on the epilepsy patient discourse .
For instance, the backbone of the \textit{Reddit} r-Epilepsy KG (Figure \ref{fig:SocialMediaKGs}, bottom, left) consists of only 17\% of the 17,558 original edges, characterizing the co-occurrence of pairs of dictionary terms in $\approx 220K$ posts by $\approx 6K$  users who posted at least once about a drug used to treat epilepsy (/S \ref{sec:data-federation}).
Due to the sparsification, this backbone is easier  to visually inspect, without any loss to the original shortest path distribution, thus preserving the distance associations among all terms revealed by the data, which are the most relevant for information spreading \cite{Correia:2023contact,soriano2023semi}.

Figure \ref{fig:backbone-graph}, shows the Ego network for the target term ``Cannabis'' (A). This is the subgraph to the \textit{Reddit} r/Epilepsy KG with all terms directly associated with ``Cannabis'' (co-occurring in posts), including associations with one another. The backbone of this graph (B) removes most of those associations. 
Note that no reachability or shortest path information is lost in the backbone sparsification; all nodes shown in Figure \ref{fig:backbone-graph}.A are reachable from the ``Cannabis'' node via an indirect path on the full backbone KG (Figure \ref{fig:SocialMediaKGs}) with the exact same shortest distance as in the original KG.
However, most are no longer directly connected to ``Cannabis'' as shown in Figure \ref{fig:backbone-graph}.B 
Importantly, the nodes that remain directly connected to ``Cannabis'' have a transitive relationship to this target node---i.e. the direct distance is shorter or equal than any other indirect path between them and to ``Cannabis''. In other words, their association with the target node is direct, and not correlated with a third or more terms.
Similarly, many edges between nodes disappear in the ``Cannabis'' ego-backbone, because their relationship via this target term is stronger than direct measurement of their co-occurrence.
In summary, from the perspective of shortest-paths, the dictionary terms that remain connected in the ``Cannabis' ego-backbone are the most relevant to understand how the Epilepsy digital cohort extracted from the \textit{subReddit} r/Epilepsy discusses this term.

Another example is the backbone of the EFA KG (Figure \ref{fig:SocialMediaKGs}, bottom, right), which consists of only 15.7\% of the 33,795 original edges, characterizing the co-occurrence of dictionary terms in $\approx 79K$ posts by $\approx 8.5K$  users who posted at least once about a drug used to treat epilepsy (/S \ref{sec:data-federation}).
Figure \ref{fig:backbone-graph}.C shows a subgraph of this backbone with all the nodes directly associated with two drugs known to treat epilepsy, \textit{Levetiracetam} and \textit{Carbamazepine} (larger purple nodes), and frequently prescribed together in refractory (drug-resistant) epilepsy.
Several terms appear in the middle of the graph that are shared by both drugs. These include additional drugs (in red) often co-prescribed with these medications in a clinical attempt to control patient seizures, and medical terms (in blue) related to the side effects of these drugs. Some are minor, such as \textit{fatigue}, \textit{somnolence}, and \textit{dizziness}.
Other terms such as \textit{Mood swings}, \textit{aggression}, \textit{depression}, and \textit{crying} are moderate to severe side effects associated with \textit{Levetiracetam} (often main reasons patients switch from this medication) and they appear in close proximity to the drug in the graph.
\textit{Nightmares} are a common side effect of both drugs, but typically worse with \textit{Carbamazepine} and appears closer to it in the subgraph. Note there is no direct backbone connection between \textit{Nightmares} and \textit{Levetiracetam}.
Also, \textit{Carbamazepine} is not safe to take during \textit{Pregnancy} as it is associated with neural tube defects. \textit{Levetiracetam} is considered safer, therefore many women of childbearing age or who plan to become pregnant will switch medications during this time, so it is interesting that the term appears between both drugs. 
Likewise, \textit{Carbamazepine} can cause significant \textit{Weight} gain, making patients switch to \textit{Levetiracetam} because it is weight neutral.

The examples above highlight how the metric backbone of KGs can be leveraged to more clearly understand how patients discuss drugs and their side effects in a particular social medium, such as the EFA forums or \textit{Instagram}. Such network analysis can be relevant to other patients as well as biomedical researchers studying these drugs. Ego-networks and other KGs have shown to be useful in various mental health and biomedical problems \cite{perry2020introduction}, and the metric backbone sparsification facilitates such analysis as discussed above.
%

%<FROM WENDY> Carbamazepine is an anti-seizure drug but also used to treat bi-polar disorder, so I am wondering if these symptoms are overlapping because the drug may be causing them in the case of levetiracetam or treating them in the case of carbamazepine. However, since these meds are so frequently used together, I wonder if what's going on here is these side effects of levetiracetam are worse when combined with carbamazepine.

%
% Network Visualization
%
\subsection{Maps of \textit{myAURA}'s knowledge graph}
\label{sec:map-visualization}

\begin{figure*}[ht!]
    %\centerline{\includegraphics[width=18.5pc]{fig-networks.png}}

\centerline{\includegraphics[width=\textwidth]{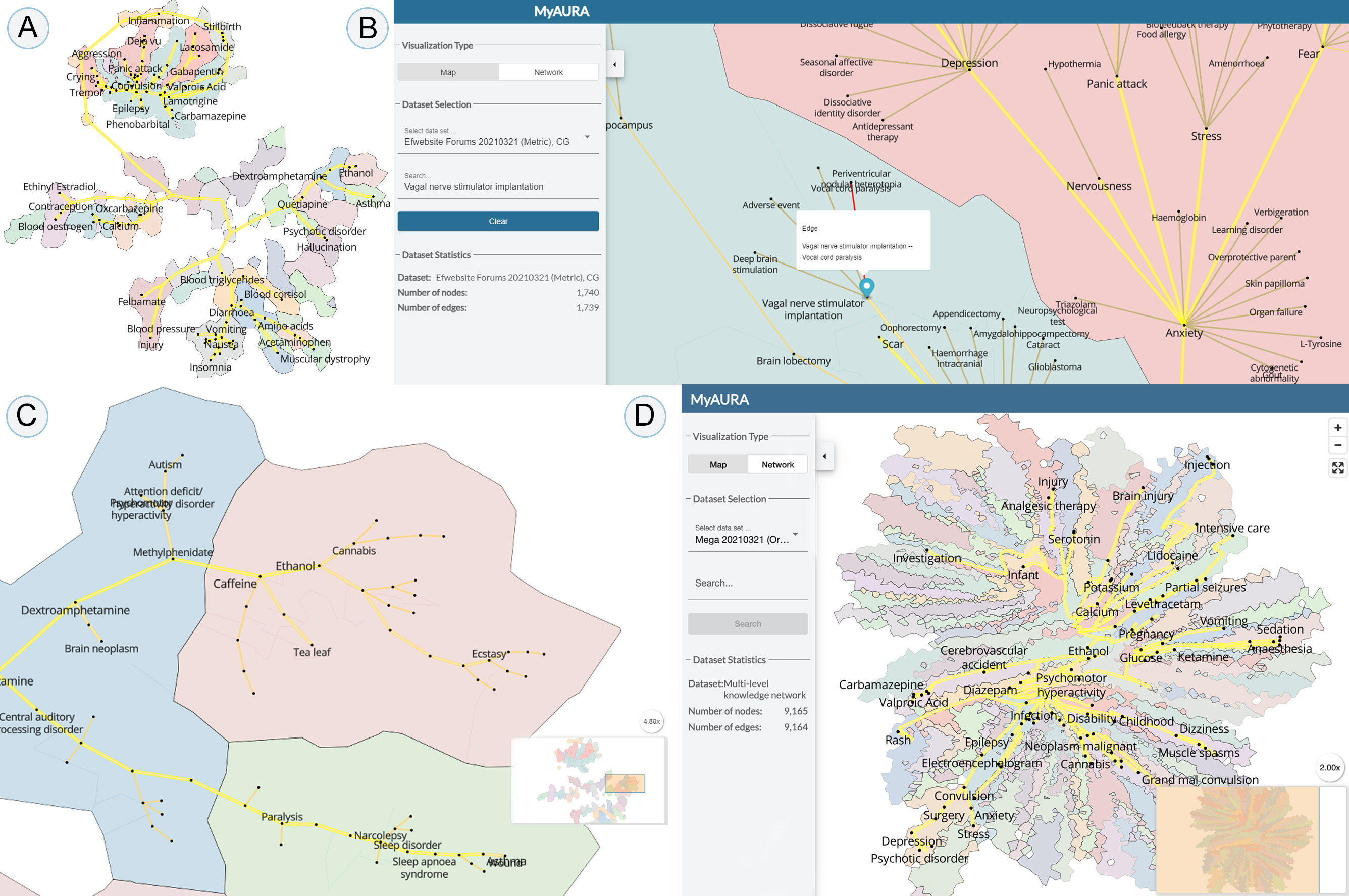}}
 %\centerline{\includegraphics[width=0.60\textwidth]{images/Figure-03-A.png}}
  %  \vspace{.3em}
    %\centerline{\includegraphics[width=0.60\textwidth]{images/Figure-03-B.png}}
    %\includegraphics[width=\textwidth]{example-image-b}
    \caption[Map Viz for myAURA KG.]{Map visualization tool for \textit{myAURA} KG  \cite{MyAuraViz}. 
    \textbf{A:} Metric backbone of EFA forums KG at the top level. \textbf{B:} \textit{myAURA} visualization tool; search and zoom in on term `Vagal Nerve Simulator Implantation', which is a ``town'' in ``surgery country'' (green), neighboring the ``anxiety/depression country''; users can click to retrieve additional information about dictionary terms related by ``roads'' (edges and paths), e.g. `Vocal cord paralysis', and the data items that co-mention both terms, such as actual EFA forum posts.
     \textbf{C:} Zoom (4.88x) on ``Ethanol country'' (pink) neighboring ``dextroamphetamine country'' (blue) and ``asthma country'' (green); the interface shows top level map (in this case the one shown in A) as an inset with the zoomed portion highlighted.
    \textbf{D:} Interface allows visualization of the combined $\varepsilon$ multi-layer KG, shown at 2x zoom level.
        This enables interactive exploration  of terms indexing data items about clinical trials, drugs, and diseases extracted from multiple sources, and their relations.
        %
        %\textbf{(top)}. Complete knowledge network shown, 2x zoom level.
        %
        %\textbf{(bottom)}. Zoomed in (5.49x) to show the epilepsy drug Diazepam and its location on the multi-layer knowledge network.
        %
        The interface can be accessed at \href{https://cns-iu.github.io/myaura}{cns-iu.github.io/myaura}  \cite{MyAuraViz}.
    }
    \label{fig:map-networks}
\end{figure*}

The full \textit{myAURA} KG, $\varepsilon$ is comprised of various networks $\{P^s (X)\}$, whose edges are extracted from distinct data sources and units of analysis (i.e., EFA comments, tweets, or paper abstracts). Thus, \textit{myAURA} users can trace the specific discourse that gave rise to an edge to understand the context in which the terms were used.
To support such analysis, we developed novel visualizations of KG backbones to leverage their interpretation in an interactive manner.
%
%Going forward, we intend to use such examples to validate the usefulness of the metric backbone of knowledge networks to epilepsy patients and researchers alike.

The inherent parsimony of associations and paths in the metric backbone of KGs makes them ideal lines of argumentation for explaining why a certain inference is made. 
Thus, we developed a \textit{myAURA} KG visualization tool \cite{MyAuraViz} using the \textit{map4sci} visualization suite \cite{ginda2023introducing,map4sci}. 
Our KG visualization differs from traditional node and edge representations of networks. Using the \textit{Zoomable Multi-Level Tree} (ZMLT) algorithm \cite{DeLuca:2019,ahmed2023splitting}, it 
charts the knowledge embedded on backbone subgraphs obtained from our various data sources onto a 2D plane resembling a cartographic map, with three graph layouts: \textit{BatchTree}, which optimizes for scalability using \texttt{C++} and \texttt{OpenMP}, balancing between a compact layout and edge length preservation; \textit{CG}, which optimizes compactness at the expense of preserving edge length; and \textit{DELG}, which optimizes to preserve edge length.

All variations are based on the metric backbone of KGs and use the same visual metaphor that displays 
%
%ZMLT uses a visual metaphor that displays 
semantic countries (defined by clusters of related dictionary terms) as regions with cities (the terms) linked by roads (the most important associations for information transmission).
We have shown with human subject studies that such map-like visualizations are as good or better than standard node-edge representations of graphs, in terms of task performance, and memorization and recall of the underlying data \cite{Saket:2015}.
Notice that semantic countries are mostly unaffected by sparsification because the metric backbone preserves community structure \cite{Correia:2023contact}.

As a user zooms in, edges down the hierarchy of importance are revealed as peripheral roads between lower importance dictionary term nodes. The tool allows easy dictionary-term search in the map, e.g. searching for the term ``Vagal nerve stimulator implantation'' as depicted in Figure \ref{fig:map-networks}.B.
The online version of the map visualization tool 
also allows clicking on edges to retrieve information associated with the connected terms.
However, due to privacy and access rights for each data source, it does not retrieve the actual data items where the terms are co-mentioned  e.g. EFA Forum posts associated with an edge connecting ``Vagal nerve stimulator implantation'' to ``Vocal cord paralysis'' as depicted in Figure \ref{fig:map-networks}.B.
Only our private, PHI-compliant research prototype is able to retrieve ranked data from all included resources after clicking on specific nodes and edges (e.g. EHR, clinical trials, or social media posts).

Importantly, the visualization allows us to represent 
\textit{myAURA}'s multi-layer KG, $\varepsilon$, as a single, two-dimensional map. 
This is done by combining edges from each constituent network $\{P^s (X)\}$ according to a specific aggregation operation \cite{Simas:2021}. In the current implementation $p_{xy}$ values from each layer are averaged across data sources, but other aggregations are possible, e.g. choosing the maximum $p_{xy}$ (minimum $d_{xy}$, see eqs. \ref{eq:proximity} and \ref{eq:distance}) in all layers as we have done in the aggregation of multi-layer protein-protein networks in another setting \cite{correia2024conserved}.
The ability to search and explore \textit{myAURA}'s heterogeneous data sources via a single combined map is a useful feature of this visualization approach as depicted in 
Figure \ref{fig:map-networks}.D.

\subsection{User-centered design and pilot testing of \textit{myAURA} through focus groups}
\label{sec:user-interface}

\begin{figure*}[ht!]
    \centerline{\includegraphics[width=0.85\textwidth]{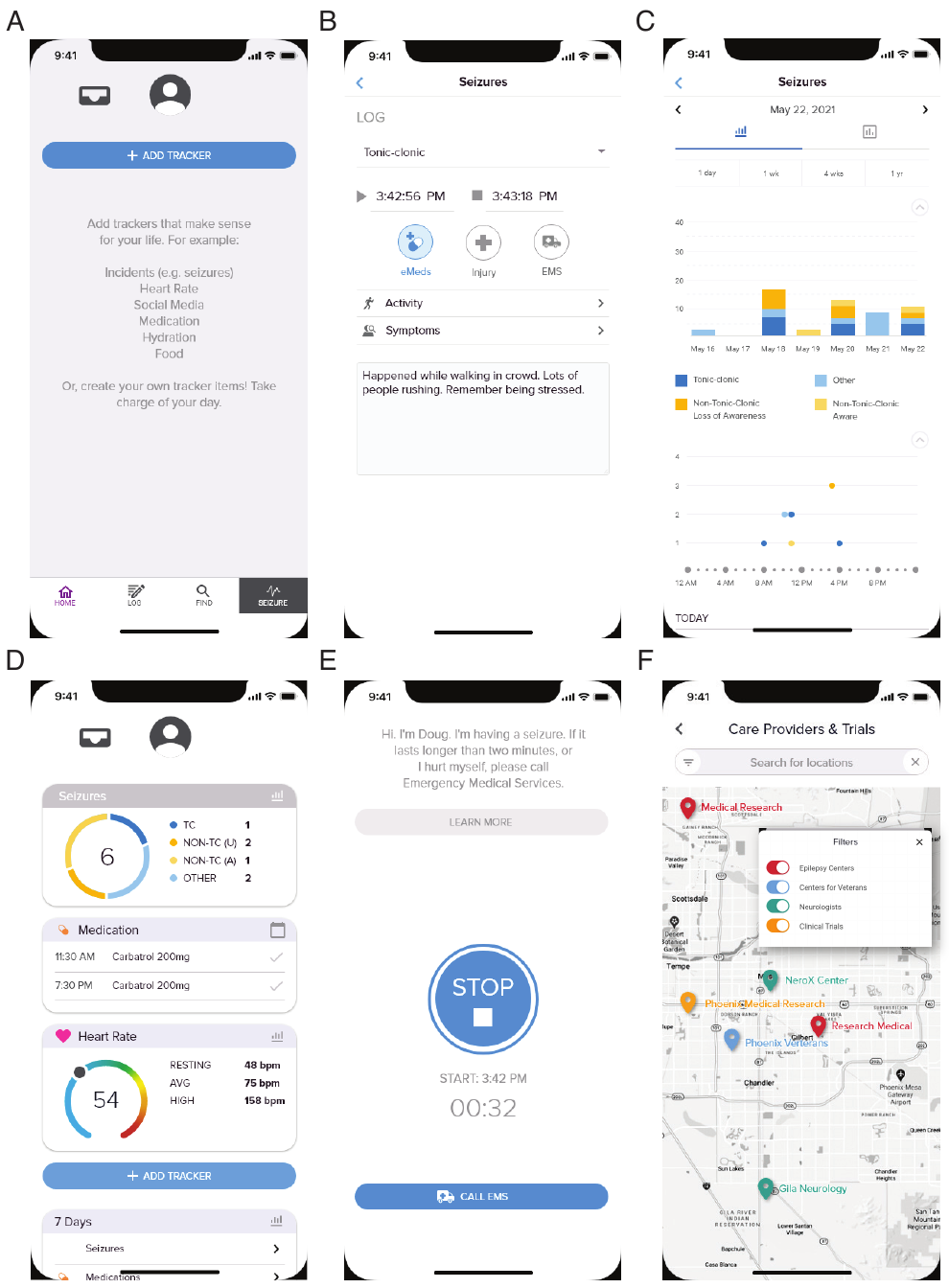}}
    \caption[]{
        The latest user interface (UI) of the \emph{myAURA} application.
        \textbf{(A)} A selection of items (e.g., seizure, medication, food), users can track.
        \textbf{(B)} Seizure log data input screen.
        \textbf{(C)} Data visualization for different types of seizure log.
        \textbf{(D)} A customizable dashboard of what is most important to the user, from seizure statistics to medication alerts and current heart rate.
        \textbf{(E)} An automatic message alert system that notifies emergency contacts in the event of prolonged seizures.
        \textbf{(F)} Map search results interface, where epilepsy centers, veteran centers, neurologists, and clinical trials can be found and filtered.
    }
    \label{fig:app-user-interface}
\end{figure*}

Our priority was to understand and include the needs of PWEC in prototyping an application to support epilepsy self-management. To best design myAURA, we carried out a series of focus group interviews to understand how to deliver personalized recommendation and visualization of information from \textit{myAURA}'s KG. The initial focus group had 12 PWEC participants that met for four sessions. 
We learnt that they experienced difficulties in  finding the right information due to diverse symptoms among PWE, as well as in tracking and managing epilepsy-related information since it is gathered via multiple sources, forcing them to use multiple apps and strategies for those goals.
They also reported difficulties in sharing information with doctors and family members and in getting support while and after having seizures.  \cite{Min:2021, Min2023CHI}.

Participants were eager for an application like \textit{myAURA} to be an epilepsy-specific, all-in-one platform to track symptoms, seizures, available treatments, and other relevant factors, and provided them and caretakers a holistic image of their epilepsy status \cite{Min:2021,Min2023CHI}. 
The ability to tailor information (e.g., finding the most effective treatments regarding individual PWE's symptoms and contexts) was also very important for them, as was the ability to share information easily with family members, friends, teachers, and health care providers. 

%

%
% YEAR 2 REPORT
%

Using these initial findings, we produced an initial interactive mockup prototype with the desired key features, and subsequently tested it with a second focus group (a subset of the original group).
Access to the initial mockup was provided at a virtual meeting where participants were asked to perform a few tasks while the screen was shared with the researcher.
They also used the mockup freely for several minutes before sharing their experience, including challenges, in a short follow-up interview.
Finally, they completed and additional survey designed to measure their perceptions and experiences with the mockup on a 7-point Likert scale \cite{Laugwitz:2008}.
Overall, the interview and survey results showed that their perceptions and experiences were positive. 
Higher scoring items were easiness to learn (6.38), feeling of control (6.19), and overall impression (6.17).
Although creativity (5.19), usefulness (5.34), and satisfaction (5.36) scored relatively lower than other items, their overall scores were still positive.

Elicited additional desired functions, they suggested items such as water/food intake trackers, medication or appointment reminders, and ability to share data with other care stakeholders (e.g., a physician, or a child’s teacher).
The ability to track and graph multiple aspects of life was deemed fundamental to provide meaningful information to PWEC, family and the medical team (e.g. identifying seizure triggers). 
%

%
% YEAR 3 REPORT
%
Based on this feedback, we designed and implemented a final interactive mockup myAURA app (see Figure \ref{fig:app-user-interface}), which included trackers (e.g., food/water intake, sleep, menstrual cycles), modifications to the navigation of screens, a dedicated media library where users may curate epilepsy-related information, an emergency/seizure response function, the ability to sync the platform with fitness trackers (e.g., FitBit), and appointment/medication reminders.

To identify additional potential functions and to understand seizure management experience in diverse environments (e.g., home, school/workplace, public transport), we conducted a (third) follow-up study with the updated mockup \cite{Min:2021}.
Our aim was to better understand the contexts, challenges, and coping strategies for seizure management devised by PWEC.
We focused on understanding the social stigma experienced by PWE and proposed human-computer interaction design requirements to effectively deliver appropriate first aid information to bystanders to a seizure \cite{Min:2021}.

The three studies further allowed us to complete a system design framework to characterize challenges PWEC face in finding the right just-in-time information, tracking, and sharing it with family, caregivers, and others. 
With this human-centered approach, we proposed a design framework to mitigate the challenges PWEC face and improve epilepsy information management and care coordination in \textit{myAURA} or a similar future technology \cite{Min2023CHI}. 
%

%
% Conclusion
%
\section{Discussion}
\label{sec:discussion}

Our interdisciplinary efforts toward building \textit{myAURA}, a personalized, easy-to-use web service for PWEC are ongoing.
Most of our efforts so far have been in researching novel data and network science methods to design and implement the computational architecture of \textit{myAURA} as a user-friendly web service to improve patient activation.
This has been translated into several important novel developments discussed next.

Our approach rests on an unprecedented collection of large-scale heterogeneous data resources of relevance to study the specific biomedical and social complexity of epilepsy, in support of PWEC, including social media and community websites, electronic health records (EHR), and biomedical databases (\S \ref{sec:data-federation}).
To integrate all that data, we developed a generalizable methodology to compute a multi-layer KG (\S \ref{sec:knowledge-graphs}), based on the federation of the constituent heterogeneous data sources (\S \ref{sec:data-federation}) in separate layers linked via the terms of a human-centered biomedical dictionary (\S \ref{sec:dictionary}). 
The power of this KG approach was exemplified in the study of the drug-drug interaction phenomenon (\S \ref{sec:KG_DDI}) in EHR \cite{Correia:2019:npj,sanchez2024prevalence}, the scientific literature\cite{zhang2022translational,wang:2024}, and social media \cite{Correia:2020:review,Correia:2016}. 

To analyze the multi-layer KG, we developed a network sparsification method (with corresponding open-source code) that allows us to extract the \textit{metric backbone} of KGs, removing edges redundant for shortest paths. It outperforms existing network sparsification methods (features (a) to (f) in \S \ref{sec:KG_backbone}) and uncovers the most important edges and pathways for inference, recommendation, and visualization \cite{Simas:2021,Correia:2023contact,soriano2023semi}. In addition to those powerful and general benefits, we showed that metric backbones of KGs reveal how patients discuss disease factors and pharmacology on social media \ref{sec:backbone_inference}, and led to another novel method to extract focused digital cohorts from general-purpose social media whose discourse is more relevant to epilepsy or other biomedical problems \cite{guo:2024}.

The metric backbone is particularly amenable to simplifying the visualization of network data \cite{Simas:2021}. Thus, we developed geospatial map-like visualizations of sparsified KGs, which enable the intuitive exploration of networks \cite{DeLuca:2019,Saket:2015}, interactive search and extraction of relevant underlying data items, and merging of \textit{myAURA}'s multi-layer KG into a single map (\S \ref{sec:map-visualization}). The ability to search and explore
\textit{myAURA}’s heterogeneous data sources via a single sparsified and combined map is a useful feature for integrating all relevant information for PWEC and epilepsy researchers.

For PWEC, the KG maps, or even the underlying sparsified network directly, can provide meaningful information in an easily consumable visual format. They can, for example, query multiple different sources of information about topics such as medications, side effects, scientific literature, and clinical trials all on one platform. The resulting visualization can display the relationships among these important topics acquired from a robust combination of data sources that PWEC would not typically be able to access, such as large samples of social media and discussion forums related to epilepsy, clinical trials, or scientific literature. 

Clinicians such as neurologists, epileptologists, nurse practitioners, physician assistants, psychologists, etc., can quickly and easily visualize knowledge about practice-relevant topics affecting PWE. 
Once a working tool is produced addressing the privacy and access rights for each data source,
the KG visualization of EHR and scientific literature data could be used quickly, even during a patient encounter, to guide assessments or recommendations in the patient’s treatment.  In particular, the access and visualization of a combination of data from social media and EHR can reveal the relationship between important issues as discussed by PWEC and in association with their health records. 
Thus, in forthcoming work, we will validate these  sparsified visualizations and different methods of combining multi-layer edges with both PWEC and epilepsy researchers.

All steps of our approach relied on on stakeholder input, whereby \textit{myAURA}'s functionalities and interface were developed and pilot tested with patient-centered design principles based on focus groups studies (\S \ref{sec:user-interface}). 
The participation of the EFA alongside the focus groups in all aims was instrumental in informing the user-centered design and development of the overall \textit{myAURA} project according to stakeholder needs. 
This includes studying how social media can assist in predicting epilepsy outcomes \cite{Correia:2020:review}, human-centered dictionary refinement \cite{Min:2023}, human-centered app design \cite{Min2023CHI,Min:2021}, epilepsy-focused digital cohort extraction \cite{guo:2024}, and our biomedical data science approach at large. To our knowledge, our team is the first to investigate PWEC practices and preferences for seeking out and curating epilepsy-related content.  
The focus on stakeholders further resulted in a design framework devloped from up-close and personal descriptions of the challenges faced by PWEC which will be leveraged to improve \text{myAURA}, and is also useful for others interested in developing a similar app \cite{Min2023CHI,Min:2021}.
Indeed, the methods we detail here, and several of the data sources we have federated (e.g., EHR, social media, clinical trials) are relevant not only to epilepsy patients but also to those with other chronic conditions.

Now that the data federation, KG construction, inference based on metric backbone sparsification, multi-layer map visualization, human-centered design requirements and pilot testing for \textit{myAURA} have been completed---with constituent methods, tools, and code shared with relevant communities---app production and deployment will continue in partnership with the EFA and other stakeholders.

\section{Conclusion}
\label{sec:conclusion}

Chronic health conditions unfold as a complex interplay among biological, psychological, and societal factors that change over time. Such complex multi-layer dynamics of human health require new science, new tools, and new interdisciplinary thinking to accelerate data-driven discovery and management of chronic conditions \cite{Pescosolido:2016,trochim2006practical,rusoja2018thinking}.
We reported the advances our team has made in developing \textit{myAURA} a personal library application prototype and suite of methods to support epilepsy research and self-management through the daunting array of treatments, drugs, interactions and side effects, diet, lifestyle, and stigma.
We worked with PWEC and stakeholders to design and pilot-test the approach, which entailed federating many large-scale heterogeneous data streams into an epilepsy knowledge graph that we analyzed using novel network inference, sparsification, and visualization methods in support of personalized recommendation, digital cohort identification, understanding of pharmacology in epilepsy, etc. 
We showed that significant advances empowered by biomedical informatics are within reach for self-management and scientific discovery in epilepsy, especially by leveraging unconventional data from EHR, social media, and digital cohorts, as well as computational and theoretical advances in characterizing and visualizing multilayer complex networks. 
We look forward to continue developing the \textit{myAURA} system towards production and deployment of a full application for epilepsy, as well expanding it to include a broad range of chronic conditions and benefit many more patients in the future.

\section{Acknowledgements}

All authors were partially funded by NIH, National Library of Medicine Grant 1R01LM012832.
In addition, R.B.C,  was partially funded by Fundação para a Ciência e a Tecnologia (grant PTDC/MEC-AND/30221/2017).
L.M.R., K.B., and X.W. were partially funded by a  National Science Foundation Research Traineeship ``Interdisciplinary Training in Complex Networks and Systems'' Grant 1735095.
L.M.R. was also partially funded by a Fulbright Commission fellowship and by Fundação para a Ciência e a Tecnologia (grant 2022.09122.PTDC).
The funders had no role in study design, data collection and analysis, decision to publish, or preparation of the manuscript.

\bibliographystyle{ama.bst}
\bibliography{references}

%%
%% Appendix
%%
\appendix

\end{document}